\documentclass[aps,prl,twocolumn,groupedaddress]{revtex4}
\usepackage{graphicx}
\usepackage{bm}

\begin{document}
\newcommand{\eq}[1]{Eq.~(\ref{#1})}
\newcommand{\fig}[1]{Fig.~\ref{#1}}
\baselineskip 11pt

%------------------------------------------------------------------------------
%       Header
%------------------------------------------------------------------------------
\title{Tunable Kondo screening in a quantum dot device}
\author{Hiroyuki Tamura,$^{1,2}$ and Leonid I. Glazman$^3$}
\affiliation{$^1$NTT Basic Research Laboratories, NTT Corporation, Atsugi, Kanagawa, 243-0198 Japan\\
$^2$CREST-JST, 4-1-8 Honmachi, Kawaguchi, 331-0012, Japan.\\
$^3$Theoretical Physics Institute, University of Minnesota, Minneapolis, Minnesota 55455}
\date{\today}
\begin{abstract}
  We consider electron transport along a single-mode channel which is
  in contact, via tunnel junctions in its walls, with two quantum
  dots. Electron tunneling to and from the dots contributes to the
  electron backscattering, and thus modifies the channel conductance.
  If the dots carry spin, the channel conductance becomes
  temperature dependent due to the Kondo effect. The two-dot device
  geometry allows for a formation of $S=1$ localized spin due to the
  indirect exchange interaction, called Ruderman-Kittel-Kasuya-Yosida
  interaction. This device offers a possibility to study the
  crossover between fully screened and under-screened Kondo impurity.
  We investigate the manifestation of such crossover in the channel
  conductance.
\end{abstract}
\maketitle

%%%%%%%%%%%%%%%%%%%%%%%%%%%%%%
%       Introduction
%%%%%%%%%%%%%%%%%%%%%%%%%%%%%%
Exchange interaction of the localized spins with the conduction
electrons affects the electron transport. At low temperatures, the
localized magnetic moment tends to be screened by the spins of the
itinerant electrons. The screening manifests itself in thermodynamic
characteristics and, more importantly, in conduction of a metal with
magnetic impurities, giving rise to the Kondo
effect~\cite{Kondo64ProgTheorPhys}.  Depending on the value of
impurity spin and on the hybridization of the localized and extended
electron states, the localized spin may be screened partially or
completely, or even over-screened by itinerant electrons. It is hardly
possible to control the spin screening for an impurity in a metallic
host matrix. However, quantum dot (QD) devices have allowed for a progress
in this direction.

The high-spin or multi-channel Kondo effect in quantum dots has been
studied both theoretically and
experimentally~\cite{InoshitaPRB93,EtoPRL00,KikoinPRL01,PustilnikPRL00,PustilnikPRB01,PustilnikPRL01,HofstetterPRL02,SasakiNature00,NygardNature00,SchmidPRL00,vanderWielPRL02,VojtaPRB02,StefaniskiActaPhysicaPolonica03}.
It has been shown that the high spin induced by the intra-dot exchange
interaction plays an important role in the Kondo physics. Recent
idea~\cite{Oreg03PRL} of a very asymmetric double-dot device proposes
to study the Kondo effect around the quantum critical point
corresponding to an over-screened localized spin. In this paper, we
consider a double-dot device suitable for the investigation of an
under-screened $S=1$ state.

The double-dot system is especially interesting because it allows for
a formation of $S=1$ localized spin due to the
Ruderman-Kittel-Kasuya-Yosida (RKKY) indirect exchange
interaction~\cite{Ruderman54,Kasuya56,Yosida57}. The RKKY interaction
between spins in QDs has been
discussed~\cite{Piermarocchi02PRL,Utsumi04PRB,Tamura04JJAP}.
Recently, an experimental evidence of the RKKY interaction in coupled
dot system has been reported~\cite{CraigScience04}.  In that
experiment, two local spins in small peripheral dots were interacting
with each other by the RKKY interaction mediated by conduction
electrons in the large central dot.
The sign of the RKKY interaction depends on the specific structure of
the electron wave functions in the central
dot~\cite{VavilovCcond-mat04,SimonCond-mat04,UsajCond-mat04}, and is
hard to predict in advance.

Here we consider a two-dot device geometry where two QDs are connected
to different sides of a single-channel quantum wire.  In this device, the exchange
couplings can be continuously modulated by applying magnetic fields.
We consider the evolution of the localized spin system between two
different screened states. In the course of the evolution, the
system passes through a special point where the local spin is
under-screened. 
The advantage of the suggested geometry is a more robust
under-screened state than in the previously considered single-dot
devices~\cite{PustilnikPRL00,PustilnikPRL01,PustilnikPRB01,HofstetterPRL02,vanderWielPRL02,Kogan03PRB}
and an easier control over the crossover to a fully screened state.
Unlike the conventional quantum dot configuration, for the
side-coupling channel geometry the Kondo effect appears as an
anomalously strong backscattering (rather than transmission) \cite{KangPRB01,KobayashiPRL04}.

\begin{figure}[b]
%\begin{center}\leavevmode
\includegraphics[scale=0.7]{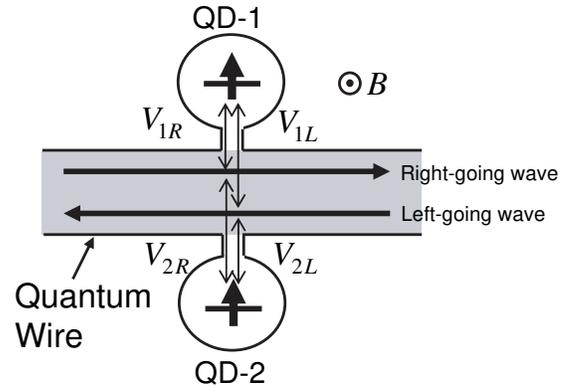}
\caption{Quantum wire coupled with two quantum dots. Two quantum dots (QD-1 and -2) coupled to a quantum wire. $V_{nk}$ is the coupling constants between dot $n$=1, 2 and the left- or right-moving waves $k=L,R$. }
\label{SchematicQD}
%\end{center}
\end{figure}

%%%%%%%%%%%%%%%%%%%%%%%%%%%%%%
%       Section
%%%%%%%%%%%%%%%%%%%%%%%%%%%%%%
% \section{Model Hamiltonian}
The Kondo interaction between a single-mode quantum wire and two QDs
($n=1,2$) each having spin 1/2 , schematically shown in
Fig.~\ref{SchematicQD}, is described by Hamiltonian
\begin{eqnarray}
&H_{\rm{ex}}=&\sum_{n kk'\sigma\sigma'}{J_{n}(k,k')c_{k'\sigma '}^{\dagger}{\bm{\sigma}}_{\sigma '\sigma}c_{k\sigma}}\cdot \bm{S}_{n},
\end{eqnarray}
where $J_{n}(k,k')=-V_{n k'}^*V_{n
  k}U_{n}/\{(U_{n}+\epsilon_{n})\epsilon_{n}\}$ and $\bm{S}_{n}$ is
the local spin at dot $n$ which has charging energy $U_n$ and
single-particle energy $\epsilon_n$.  The coupling strength is given
by $V_{n k}=v_{nk}\exp(ikx_{n})$ where the lateral coupling $v_{nk}$
depends on the overlap of wavefunctions between the QD and the wire in
the lateral direction. Two local spins in two dots interact with each
other by the RKKY interaction,
$H_{\mathrm{RKKY}}=-J_{\mathrm{RKKY}}(\bm{S}_1\cdot
\bm{S}_2)$, where $J_{\rm{RKKY}}(R)=-\pi E_F|\rho
J_1 \rho J_2| \mathrm{si}(2k_FR)$ and $J_n=|J_n(k_F,-k_F)|$; function
$\mathrm{si}(x)=-\int_x^{\infty} d\tau(\sin \tau/\tau)$ is the sine integral
function \cite{Tamura04JJAP}.

%%%%%%%%%%%%%%%%%%%%%%%%%%%%%%
%       Section
%%%%%%%%%%%%%%%%%%%%%%%%%%%%%%
%\section{Diagonalization of the $S=1$ Kondo term}
When two dots are on the opposite sides with zero longitudinal
distance $R=x_1-x_2=0$ between them, the RKKY interaction is
ferromagnetic and the exchange coupling constant is maximal,
$J_{\mathrm{RKKY}}\sim E_F(\rho J)^2$. In the case of ferromagnetic
interaction, the total spin $\bm{S}=\bm{S}_1+\bm{S}_2$ of the two dots
is $S=1$, and within the triplet sub-space of states $|t\rangle$ the
spin difference $\langle t_1|(\bm{S}_1-\bm{S}_2)|t_2\rangle=0$. The
effective Hamiltonian within that subspace is
\begin{eqnarray}
&H_{\rm{ex}}&=\bm{S} \cdot \frac{1}{L} \sum_{kk'\sigma\sigma'} \left({I^{RR} c_{Rk'\sigma '}^{\dagger}{\bm{\sigma}}_{\sigma '\sigma}c_{Rk\sigma}}\right.\nonumber\\
&&+I^{RL} c_{Rk'\sigma '}^{\dagger}{\bm{\sigma}}_{\sigma '\sigma}c_{Lk\sigma}
+I^{LR} c_{Lk'\sigma '}^{\dagger}{\bm{\sigma}}_{\sigma '\sigma}c_{Rk\sigma}\nonumber\\
&&\left.{+I^{LL} c_{Lk'\sigma '}^{\dagger}{\bm{\sigma}}_{\sigma
      '\sigma}c_{Lk\sigma}}\right),
\label{h-eff}
\end{eqnarray}
with the exchange constants
\begin{eqnarray}
&I^{RR}&=I(k_F,k_F),\ I^{LL}=I(-k_F,-k_F),\\
&I^{RL}&=I(-k_F,k_F)=(I^{LR})^*,
\label{i-lr}
\end{eqnarray}
where
\begin{equation}
2I(k,k')=J_1(k,k')e^{i(k'-k)R/2}
+iJ_2(k,k') e^{i(k-k')R/2} .
\label{j-12} 
\end{equation}
The $2\times 2$ exchange term in Eq.~(\ref{h-eff}) can be easily
diagonalized by the unitary transformation of the basis
$(c_{Rk\sigma},c_{Lk\sigma})$; the result is
\begin{eqnarray}
\lefteqn{
H_{\rm{ex}}=\bm{S} \cdot \frac{1}{L} \sum_{kk'\sigma\sigma'}\left({J_{a} a_{k'\sigma '}^{\dagger}{\bm{\sigma}}_{\sigma '\sigma}a_{k\sigma}} \right.}\hspace{1cm}\nonumber\\
&\left.{+ J_{b} b_{k'\sigma '}^{\dagger}{\bm{\sigma}}_{\sigma '\sigma}b_{k\sigma}}\right),\nonumber\\
\lefteqn{
J_{a(b)}=\frac{1}{2}\left\{I_+^{RR}+I_+^{LL}\right.}\hspace{1cm}\nonumber\\
&\pm \left[{(I_+^{RR}-I_+^{LL})^2+4(I_+^{RL})^2\cos^2 k_FR }\right.\nonumber\\
&\left.{\left.{+ 4(I_-^{RL})^2\sin^2 k_FR}\right.]^{1/2}}\right\},
\label{h-ab}
\end{eqnarray}
where $2I_{\pm}^{RR}=J_1(k_F,k_F)\pm J_2(k_F,k_F),\ 
2I_{\pm}^{RL}=J_1(k_F,-k_F)\pm J_2(k_F,-k_F),\ 
2I_{\pm}^{LR}=J_1(-k_F,k_F)\pm J_2(-k_F,k_F)$, and 
$2I_{\pm}^{LL}=J_1(-k_F,-k_F)\pm J_2(-k_F,-k_F)$.  For symmetrically
positioned dots ($R=0$) and at $B=0$, the exchange constants are
$J_{a}=J_1+J_2$ and $J_{b}=0$. It means that one of the channels is
decoupled from the spin and becomes free propagating. Note that the
decoupling for the symmetric case at $B=0$ always occurs, {\it i.e.}, 
regardless of dot parameters ($U_n,\ \epsilon_n, V_n$, etc.).

The basis $(a_{k\sigma},b_{k\sigma})$ is related to the original
one, $(c_{Rk\sigma},c_{Lk\sigma})$, by the unitary transformation:
\begin{eqnarray}
%\lefteqn{
&&\pmatrix{c_{Rk\sigma}\cr c_{Lk\sigma}}=\pmatrix{\cos\frac{\theta}{2}e^{-i\gamma /2} & -\sin\frac{\theta}{2}e^{-i\gamma /2} \cr
\sin\frac{\theta}{2}e^{i\gamma /2} & \cos\frac{\theta}{2}e^{i\gamma /2}}\pmatrix{a_{k\sigma} \cr b_{k\sigma}},
\label{UnitaryTransformation}\\
&&\cot\theta=\frac{I_+^{RR}-I_+^{LL}}{2\sqrt{(I_+^{RL}\cos k_FR)^2 +
    (I_-^{RL}\sin k_FR)^2}},
\label{theta}\\
&&e^{i\gamma}=
\frac{I_+^{RL}\cos k_FR +iI_-^{RL}\sin k_FR}{\sqrt{(I_+^{RL}\cos k_FR)^2 + (I_-^{RL}\sin k_FR)^2}}.
\end{eqnarray}

In order to evaluate conductance of the device, we start with the
definition of the operator of ``backward current''\cite{Kane92}. The presence of backward current
reduces the conductance from its maximal value (achieved for the
free-propagating modes),
\begin{eqnarray}
&I_{\textrm{back}}&=\frac{e}{2}\frac{d}{dt}\sum_{k\sigma}(c_{Rk\sigma}^{\dagger}c_{Rk\sigma}-c_{Lk\sigma}^{\dagger}c_{Lk\sigma})\nonumber\\
&&=-\frac{e}{2}\sin\theta \frac{d}{dt}\sum_{k\sigma}(a_{k\sigma}^{\dagger}b_{k\sigma}+b_{k\sigma}^{\dagger}a_{k\sigma}),
\end{eqnarray}  
where we have used a relation
$d\{a(t)^{\dagger}a(t)\}/dt=d\{b(t)^{\dagger}b(t)\}/dt=0$ for the
decoupled Hamiltonian $H=H_{a}+H_{b}$. Having the definition of
$I_{\textrm{back}}$, we may proceed with the evaluation of the
corresponding ``backward conductance'' $G_{\rm back}$ in known
ways~\cite{PustilnikPRL01,PustilnikCondmat2005}.
The conductance of the device is then calculated as
\begin{equation}
G=\frac{2e^2}{h}-G_{\rm back}.
\label{G}
\end{equation}
Note that in the derivation we assumed that the tunnel coupling of the
dots to the channel is weak; the device conductance may be cast
in the form of Eq.~(\ref{G}) only under that condition.

In the symmetric case ($J_b=0$), Kondo correlations develop only in
the channel $a$ of the Hamiltonian Eq.~(\ref{h-ab}). This point in the
parameter space corresponds to the under-screened $S=1$ Kondo spin. We
define the corresponding Kondo temperature as
\begin{equation}
T_{a}=D\exp(-1/\rho J_a),
\label{ta}
\end{equation}
where $\rho$ is the density of states of the itinerant fermions at the
Fermi level. In the perturbative regime, $T\gg T_a$, the conductance
evaluated in the leading-logarithmic approximation is
\begin{eqnarray}
G=\frac{2e^2}{h}-G_0\frac{\pi^2}{2}\frac{1}{\{\ln (T/T_{a})\}^2},
\label{g-a1}
\end{eqnarray}
where the conductance factor $G_0=(2e^2/h) \sin^2\theta$ depends on
the specific values of the exchange constants, see Eq.~(\ref{theta}).

\begin{figure}[t]
%\vspace{0.0in}
%\begin{center}\leavevmode
\includegraphics[scale=0.65]{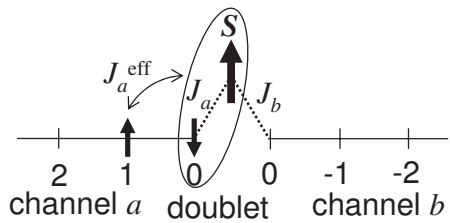}
\caption{Kondo model on lattice. Below the Kondo temperature $T_{a}$, a conduction electron $\bm{s}_{a0}$ at $n=0$ and a local spin $\bm{S}$ form a doublet state $\bm{S}'$. Then, the effective coupling between an electron spin $\bm{s}_{a1}$ on site $n=1$ and the doublet spin $\bm{S}'$ is ferromagnetic $J_{a}^{\rm eff}<0$ .}
\label{2-channelLatticeKondoModel}
%\end{center}
\end{figure}

At low temperature $T\ll T_{a}$, electrons in the channel $a$ screen
the local spin $S=1$ thus reducing it to $S=1/2$. The existence of the
residual interaction with the partially screened localized spin leads
to the logarithmic corrections in the conductance at low temperatures
too~\cite{PustilnikPRL01,PosazhennikovaPRL05}. Here we adapt the ideas
developed in Ref.~\cite{NB80} for the description of under-screened
magnetic impurity in order to derive the effective low-energy
Hamiltonian of a channel with two attached dots forming a localized
spin $S=1$. To derive the effective Hamiltonian, we consider Kondo
problem with $S=1$ on a one-dimensional lattice,
$
%\begin{equation}
\mathcal{H}=-\sum_n t_0(a_{n+1}^{\dagger}a_n 
+ \textrm{c.c})+2J\bm{s}_{a0}\cdot \bm{S}
%\label{model}
%\end{equation}
$, 
where $a_n (a_n^{\dagger})$ is the annihilation (creation) operator
for the channel $a$ on the site $n$ and $t_0>0$ is the transfer
integral between neighboring sites.  When $J>0$ two spins
$\bm{s}_{a0}$ and $\bm{S}$ tend to form a doublet. To describe the
low-temperature limit behavior of the system, the parameters of the
effective Hamiltonian 
%(\ref{model}) 
must be tuned to satisfy the
condition $|J|\gg t_0$, see Ref.~\cite{W75}. In this case the doublet
states are
$|\psi_{\frac{1}{2}}\rangle=\sqrt{2/3}|-\frac{1}{2},1\rangle -
\sqrt{1/3}|\frac{1}{2},0\rangle$, and
$|\psi_{-\frac{1}{2}}\rangle=-\sqrt{2/3}|\frac{1}{2},-1\rangle +
\sqrt{1/3}|-\frac{1}{2},0\rangle,$ where we have denoted the spin
state as $|s_zS_z\rangle$ with $s_z=\pm 1/2$ and $S_z=1,0,-1$ (see
Fig.~\ref{2-channelLatticeKondoModel}).  This doublet is the effective
spin $1/2$ formed as the result of screening at $T\lesssim T_a$.
By doing the second-order perturbation theory in the transfer
amplitude $t_0(\ll J)$ between the sites $n=0$ and $1$, we obtain
Kondo Hamiltonian for the effective spin~\cite{NB80,NozieresJP80}
coupled ferromagnetically to the rest of the system,
$\mathcal{H}_{a}^{\rm eff}=-\sum_n t_0(a_{n+1}^{\dagger}a_n +
\textrm{c.c}) + 2J_{a}^{\rm eff}\bm{s}_{a1}\cdot \bm{S}'$, where
$J_{a}^{\rm eff}\propto -t_0^2/JS<0$ and $S'=S-1/2$.

We may perform the poor man's scaling procedure then, to
obtain the effective scale-dependent interaction constant $\rho
J=1/\ln (D_0/T)$. Taking into account that the initial stage of
screening occurred at $T\sim T_a$, we have to set $D_0\simeq T_a$.
In order to evaluate conductance in the low-temperature limit, first
we consider the effect of the doublet formation on scattering. Let us
start with $S$ matrix for half-wire $n\ge 0$. The scattering state is
expressed by $\psi_{n\ge
  0}=e^{-i\frac{\pi}{2}n}+e^{2i\delta}e^{i\frac{\pi}{2}n}$.
At $T=0$, the scattering phase shift is $\delta=\pi/2$, so
$\psi\propto \sin(\pi n/2)$. If we define now the scattering matrix
for the half-chain $n\ge 1$, thus excluding the site involved in the
singlet formation, the corresponding phase shift (at the Fermi level)
would be $\tilde\delta=\delta+\pi/2$. This rule for the phase shifts
tells us that small deviations from the unitary limit for the backward
conductance lead to its reduction, 
\begin{equation}
G_{\rm back}=G_0\Bigg[1-\frac{3\pi^2}{16}\frac{1}{\ln^2(T_a/T)}\Bigg].
\label{low-temp}
\end{equation}
One finds the full form of the low-temperature limit for conductance
by substituting Eq.~(\ref{low-temp}) in Eq.~(\ref{G}).

\begin{figure}[t]
\includegraphics[scale=0.6]{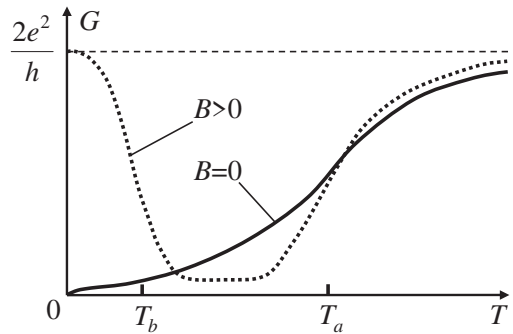}
\caption{Schematic dependence of the conductance as a function of temperatures $T$ and magnetic fields $B$.}
\label{T-dependence}
\end{figure}

Tuning of $J_b$ through the $J_b=0$ state may be achieved by applying
magnetic field to the device, which would cause the variation of the
orbital parts of electron wave functions. Now we turn to the case of
non-symmetric device, in which case $J_b\ne 0$. There are two
characteristic temperatures now, $k_BT_{\gamma}=D\exp\left(-1/2\rho
  |J_{\gamma}|\right)$, where $\gamma=a,b$. The perturbative
calculation in the leading-logarithmic approximation yields at $T\gg
T_a, T_b$,
\begin{equation}
G=\frac{2e^2}{h}-G_0\frac{\pi^2}{2}
\Bigg[\frac{1}{\ln(T/T_{a})}-\frac{1}{\ln(T/T_{b})}\Bigg]^2.
\label{big-T}
\end{equation}
For definiteness, in the following we assume $T_a\gg T_b$. In the
intermediate interval of temperatures, $T_b\ll T\ll T_a$, we again may
use an effective Kondo Hamiltonian with the residual local spin
$S=1/2$. Derivation similar to the sketched above for the case of
$T_b=0$, now yields
\begin{equation}
G_{\rm back}=G_0\left\{1-\frac{3\pi^2}{16}
\left[\frac{1}{\ln (T/T_{a})}-\frac{1}{\ln (T/T_{b})}\right]^2\right\}.
\label{inter-T}
\end{equation}
For $T\ll T_{b}\ll T_{a}$, the remaining spin 1/2 is also screened by
channel $b$, and the Nozieres Fermi liquid theory can be
applied~\cite{NozieresJP80}. By employing the calculational method
given in Ref.~\cite{PustilnikPRL00}, we find
\begin{equation}
G=\frac{2e^2}{h}-G_0\left(\frac{\pi T}{T_a}-\frac{\pi T}{T_b}\right)^2.
\label{low-T}
\end{equation}
Figure~\ref{T-dependence} schematically summarizes the temperature dependence of the conductance.

We estimate the Kondo temperatures in a
GaAs quantum wire coupled with two identical QDs at zero longitudinal
distance.  For simplicity, we assume the dot-wire coupling strength is
written as $v_{nk}=V\phi_{k_F}(y_0)$ where the coupling is determined
by the amplitude of the lateral wavefunction $\phi_k(y)$ at $y=y_0$.
Then, we obtain $J_{RKKY}=(\pi^2/2)E_F(\rho J)^2(\phi_L \phi_R)^2$ and
$\rho J_{a(b)}=\rho J(\phi_L \pm \phi_R)^2/2$ where $\rho J={2\rho
  U|V|^2}/\{\epsilon(\epsilon+U)W\}$, $W$ is the wire width and $\phi_{L(R)}\equiv
\sqrt{W/2}\phi_{k_F}(\pm y_0)$.  At $B=0$, $\phi_L=\phi_R$, and then, $J_a\simeq
J$ and $J_b=0$.  As increasing magnetic fields, $\phi_L$ rapidly
decreases to zero since it is pushed towards to the other side of
wall, whereas $\phi_R$ has only weak magnetic field dependence.  In
the high magnetic field limit, $\phi_L\rightarrow 0$ and $T_b
\rightarrow T_a$.  Although this qualitative behavior is not changed
by a detail of the type of the dot-wire coupling or the confinement
potential, to estimate a quantitative ratio of the Kondo temperatures
$T_a/T_b=\exp({1/2\rho J_a -1/2\rho J_b})$, we assume the lateral
confinement of the wire at $x=0$ is a square-well potential with width
of $W=$50 nm, and $y_0=\pm 0.3W$.  The Fermi energy is 10~meV
corresponding to the Fermi wave length of 54~nm so that the wire has
only one conducting channel.  We choose $\rho J$=0.2, which is deduced
in the typical measurement of the Kondo effect \cite{WielScience00}
giving the upper Kondo temperature $T_a\sim 1$~K.  A numerical
calculation of the single-particle wavefunction for the square-well
confinement gives $T_b\simeq 0.25T_a$ and $J_{\mathrm{RKKY}}\simeq 0.2$~
meV at $B=2.5$~T.  This evaluation confirms that the two Kondo
temperatures are in an accessible range in experiment while the RKKY
interaction is still large enough to maintain the spin triplet.

So far, we have assumed the wire has only one conducting channel where
the under-screened Kondo effect can be experimentally
realized without employing a complicated gate tuning~\cite{Sasaki05}.
It is interesting to discuss how the
Kondo screening changes when the channel number exceeds
$2S=1$. In the case of perfect alignment of the dots and at $B=0$, it
is still possible to single out one propagating mode interacting with
the local spin, so the problem still maps on a single-channel Kondo
problem ($T_a\neq 0$, $T_b=0$). At $B\neq 0$, however, a number of
modes couple to the local spin, so the intermediate-temperature
behavior of the conductance is more complicated than the one given by
Eqs.~(\ref{big-T}) and (\ref{inter-T}). However, at the lowest
temperatures two channels coupled to the spin strongest still will
lead to a full screening, see Eq.~(\ref{low-T}).

%%%%%%%%%%%%%%%%%%%%%%%%%%%%%%
%       Summary
%%%%%%%%%%%%%%%%%%%%%%%%%%%%%%
In summary, we discussed electron transport along a single-mode
channel which is in contact with two side-coupled quantum dots. If
each dot has a spin 1/2, the two-dot device geometry allows for a
formation of $S=1$ localized spin due to the indirect RKKY exchange
interaction. We investigate the temperature and magnetic-field
dependence of the conductance for such a device which shows the
crossover between fully screened and under-screened Kondo impurity.

%%%%%%%%%%%%%%%%%%%%%%%%%%%%%%
%       Acknowledgments
%%%%%%%%%%%%%%%%%%%%%%%%%%%%%%
We are grateful to M.~Pustilnik and H.~Takayanagi for valuable discussions. This work
was supported by the CREST Project of the Japan Science and Technology
Corporation (JST), and the NAREGI Nanoscience Project, Ministry of
Education, Culture, Sports, Science and Technology, and by NSF Grants
DMR02-37296 and EIA02-10736.

%%%%%%%%%%%%%%%%%%%%%%%%%%%%%%
%       References
%%%%%%%%%%%%%%%%%%%%%%%%%%%%%%

\end{document}